\documentclass[conference]{IEEEtran}
\usepackage{cite}
\usepackage{amsmath,amssymb,amsfonts}
\usepackage{graphicx}
\usepackage{textcomp}
\usepackage{xcolor}
\usepackage{booktabs} 
\usepackage{graphicx}
\graphicspath{ {figures/} }
\usepackage{algpseudocode}
\usepackage{algorithm}

\usepackage{url}

\usepackage{booktabs}
\usepackage{balance}
\begin{document}

\title{SATA$n$: Air-Gap Exfiltration Attack via Radio Signals From SATA Cables}

\author{\IEEEauthorblockN{Mordechai Guri}
	\IEEEauthorblockA{Ben-Gurion University of the Negev, Israel\\
		Department of Software and Information Systems Engineering\\ Cyber-Security Research Center \\
		Email: gurim@post.bgu.ac.il\\
	Demo video: http://www.covertchannels.com}}


\maketitle

\begin{abstract}
This paper introduces a new type of attack on isolated, air-gapped workstations. Although air-gap computers have no wireless connectivity, we show that attackers can use the SATA cable as a wireless antenna to transfer radio signals at the 6 GHz frequency band. The Serial ATA (SATA) is a bus interface widely used in modern computers and connects the host bus to mass storage devices such as hard disk drives, optical drives, and solid-state drives. The prevalence of the SATA interface makes this attack highly available to attackers in a wide range of computer systems and IT environments. We discuss related work on this topic and provide technical background. We show the design of the transmitter and receiver and present the implementation of these components. We also demonstrate the attack on different computers and provide the evaluation. The results show that attackers can use the SATA cable to transfer a brief amount of sensitive information from highly secured, air-gap computers wirelessly to a nearby receiver. Furthermore, we show that the attack can operate from user mode, is effective even from inside a Virtual Machine (VM), and can successfully work with other running workloads in the background. Finally, we discuss defense and mitigation techniques for this new air-gap attack. 

\end{abstract}

\begin{IEEEkeywords}
air-gap, network, exfiltration, electromagnetic, leakage, covert channels, SATA 
\end{IEEEkeywords}

\section{Introduction}
Information is one of the organization's valuable assets in the digital era and, accordingly, is coveted by adversaries. There are many types of threats to the organization's data, including data theft, malware, spyware, ransomware, advanced persistent threats, data leakage, data misuse, etc.   
Air-gap is a network security measure taken where highly sensitive data is involved, in order to protect the information from cyber attacks or accidental leakage. In this measure, the local area network (LAN), computerized systems, or specific device is maintained in an isolated environment, disconnected from the Internet or other non-secure networks. Air-gap policies are commonly backed up with supporting regulations, such as forbidding Wi-Fi and Bluetooth connections, restricting the use of external media, enforcing access control, and using Anti-Virus (AV) and Endpoint Detection and Response (EDR) products. The air-gapped measure is used in many IT environments, such as military and defense networks, industrial control systems, government agencies, and banking and finance sectors \cite{Guri:2018:BAM:3200906.3177230}.

\subsection{Air-Gap Penetration}
But even offline networks, which are completely disconnected from the Internet, can be hacked. It has been proven that motivated and persistent threat actors can breach the air-gap isolation, installing advanced persistent threat (APT) in the network. The most notorious example is the Stuxnet attack from 2010 designed to target nuclear facilities by targeting programmable logic controller (PLC) units \cite{langner2011stuxnet}. In this case, the attackers targeted Microsoft Windows machines and spread them through USB drives plugged into the air-gapped machines on the network. Since 2010, more than ten new APTs have been reported targeting air-gapped facilities, including ProjectSauron, EZCheese, and USBCulprit \cite{dorais2021jumping}. In 2019 security firms reported the Ramsay Advanced Persistent Threat (APT), a cyber-espionage malware that was targeting air-gapped networks. This malware moves data between isolated networks and Internet-connected computers using external USB thumb drives \cite{ESETRese49:online}. Such cases prove that the risk is increased because the air-gapped isolation brings a false sense of security that the systems are immune to breaches.

\subsection{Covert Channels}
Compromising the air-gapped network is only the first operative phase for an attacker. For espionage purposes, the APT moves to a collection phase, where it gathers different types of information; files, images, keylogging, etc. In the case of Internet-connected networks, APTs commonly use different covert communication channels to hide the data and leak it outward. Over the years, many types of covert channels have been revealed, researched, and analyzed. To evade DLP and monitoring solutions, an attacker may conceal data in ICMP, HTTP(S), DNS, SMTP, and other common protocols \cite{mazurczyk2016information,murdoch2005embedding}. However, because air-gapped networks lack connectivity to the Internet, the attacker must use non-standard ways to exfiltrate data. The air-gap covert channels explored over the years include acoustic, optical, electromagnetic, electric, and other types of physical covert communication techniques \cite{Guri:2018:BAM:3200906.3177230,carrara2016air}.



This paper presents a new type of electromagnetic air-gap covert channel. The covert enables attackers to leak data from air-gapped systems using the SATA cable as a transmitting antenna. We present the design and implementation and discuss the evaluation of the covert channel.

\subsection{Paper Organization}
This paper is organized as follows. The attack model on air-gapped networks is explained in Section \ref{sec:attack}. Related work is discussed in Section \ref{sec:related}. Technical background on the SATA interface is given in Section \ref{sec:tech}. The design, implementation, and algorithms of the transmitter and receiver are described in Section \ref{sec:trans}. Section \ref{sec:eval} presents the evaluation results. Countermeasures are proposed in Section \ref{sec:counter}. We conclude in Section \ref{sec:conclusion}.

\section{Attack Model}
\label{sec:attack}
To steal valuable assets such as sensitive information, financial data, and intellectual property, attackers may employ an offensive strategy known as the advanced persistent threat (APT).
The life cycle of a modern advanced persistent threat consists of various phases. The main steps are the initial penetration, establishing foothold, lateral movement, data collection, and exfiltration \cite{assante2015industrial}. 

\subsection{Initial Penetration}
In this phase, the attacker breaches the layers of defense and installs malware in the target network.
In the case of a standard connected network, this step is commonly performed using social engineering, spear-phishing, zero-day exploits, and malicious web pages. However, when the target network is disconnected from the Internet (air-gapped), an attacker may use complex methods such as supply chain attacks, removable media attacks, malicious insiders, and deceived employees to breach the network \cite{mcfadden2010supply,bahrami2019cyber}. 
Various advanced cyber attacks targeting air-gapped networks have been publicized since 2010, including Agent.BTZ, Stuxnet, ProjectSauron, Emotional Simian, and USBCulprit and Ramsay \cite{dorais2021jumping}. Most of these attacks use removable media, such as USB thumb drives to infect workstations within the air-gapped network. For example, the Agent.btz is a computer worm that breached U.S. military networks via an infected USB drive attached to a computer in the network . Similarly, the Stuxnet malware attacked supervisory control and data acquisition (SCADA) systems in 2010 \cite{kushner2013real}. In 2020, researchers published technical details on USBCulprit \cite{guri2021usbculprit} and Ramsay \cite{ESETRese49:online}, APTs which seems to be designed to reach air-gapped networks. 

\subsection{Data Collection}
After establishing a foothold in the target network and expanding control to other workstations, servers, and infrastructure, the attack move to the data collection phase. In this phase, various data of interest are gathered and collected. The information may vary from victim to victim and includes files with sensitive information, keylogging, emails, images, keylogging, etc. The attacker may encrypt or hide the data at this stage, concealing it from data leakage prevention (DLP) solutions \cite{mazurczyk2015information}.

\subsection{Data Exfiltration (SATAn)}
At some point, the attacker might want to exfiltrate the data. In this stage, the malware locates workstations or servers in the network that contains active SATA interfaces; computers with Hard Disk Drives (HDD), Solid State Drives (SSD), or optical drives such as CD/DVD. The malware then uses a specialized shellcode to maintain file system activity to generate radio signals from the SATA cables. The collected data is modulated, encoded, and transmitted via this covert channel.

\subsection{Data Reception}
The exfiltrated information can be received in different ways. A hardware receiver might be hidden on implanted near the air-gapped computer. For example, in 2018, tiny microchips were found hidden inside servers used by Apple, Amazon, and government contractors. According to a report by Bloomberg Businessweek, the chip allows control of the compromised computer \cite{TheBigHa0:online}. Another way is to have a malicious insider or visitor carry a radio receiver nearby the air-gapped computer, for instance, within a laptop. The receiver monitors the 6 GHz spectrum for a potential transmission, demodulates the data, decodes it, and sends it to the attacker.

\begin{figure*}
	\centering
	\includegraphics[width=0.7\linewidth]{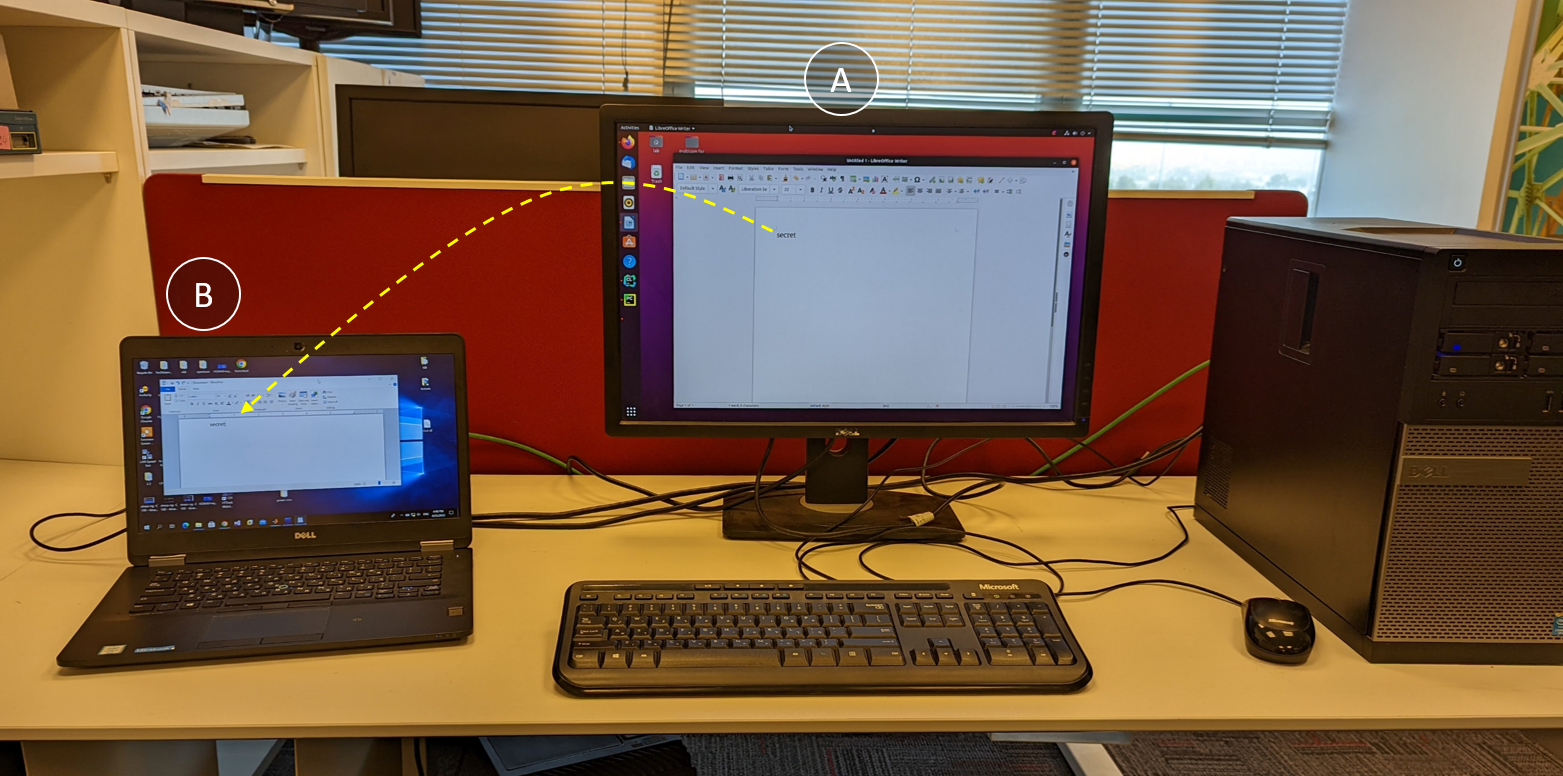}
	\caption{Demonstration of the SATA$n$ covert channel. A piece of sensitive information in an air-gapped workstation (A) is transmitted via radio signals from the SATA cable to a nearby laptop receiver (B). In this case, the word 'SECRET' was transmitted.}
	\label{fig:il}
\end{figure*}

Figure \ref{fig:il} shows the demonstration of the covert channel. Sensitive information in an air-gapped workstation (A) is transmitted via radio signals from the SATA cable to a nearby laptop receiver (B). In this case, the word `SECRET' was transmitted.

\section{Related Work}
\label{sec:related}
The covert channel term, coined in 1973 by Butler Lampson, is defined as communication channels that are not intended for information transfer \cite{lampson1974protection}. Covert communication channels have been explored for many years. Attackers may use legitimate network traffic to conceal and hide data in traditional covert channels. For instance, information may be hidden within TCP headers, HTTPS requests, DNS extra fields, and SMTP messages \cite{mazurczyk2016information}. The attacker may also use techniques such as stenography and image or video manipulations to hide textual binary data \cite{mazurczyk2015information}.

Air-gap covert channels are special types of covert communication channels that enable attackers to leak data from isolated, air-gapped systems where no standard networking exists. In this domain, the attacker uses physical mediums to modulate information into the air. These methods can be mainly classified into electromagnetic, magnetic,  electric, optical, and acoustic \cite{Guri:2018:BAM:3200906.3177230}.

Information can be leaked from network-less systems via electromagnetic waves. In this technique, malware triggers electromagnetic emissions from various computer components such as buses, cables, and processors. Data is modulated on these signals and broadcast to the environment where a potential radio receiver receives the data, process, and demodulate it. Electromagnetic-based channels have been proposed and explored in the covert channels research domain for many years. The AirHopper attack used the video cards in air-gapped computers to generate FM signals modulated with the leaked information \cite{guri2014airhopper}. The USBee attack used the A Universal Serial Bus (USB) data buses to control electromagnetic waves emanated from the USB and encode binary data over it \cite{guri2016usbee}. Other work such as GSMem \cite{guri2015gsmem}, BitJabber \cite{9300268}, EMR \cite{LORAMER}, AIR-FI \cite{guri2020air}, and CloakLoRa \cite{9259364} used the memory modules and CPU to generate radio waves from an air-gapped systems for data exfiltration. 

Although the electromagnetic-based covert channels can be mitigated with Faraday cages, researchers found a way to bypass them. Guri et al. proposed using magnetic fields to evade Faraday shields and maintaining covert communication to leak data from Faraday caged air-gapped computers to nearby magnetic sensors \cite{guri2019odini, ,GURI2021115}. PowerHammer is a cyberattack that allows data exfiltration via emission from power supplies emanated to the power lines \cite{guri2019powerhammer}. Shao et al. proposed using the noise in the power lines for communication \cite{shao2020your}. 

Information can also be leaked from air-gapped systems in optical ways. In general, optical communication is defined as any type of communication in which light is used to carry signals. In the case of covert channels, optical sources in computers are used to carry signals, which are received by remote cameras or photo-detector, which convert light into electricity using the photoelectric effect. Loughry \cite{loughry2002information} and recently Guri \cite{guri2019ctrl} proposed to use the status LEDs of workstation keyboards to encode information. Nassi et al. used lasers and scanners to infiltrate air-gapped networks in the organization \cite{nassi2018xerox}. The LEDs in network devices, cameras, and screens \cite{guri2016optical} were also used for data leakage purposes \cite{Guri2017}. Other works such as \cite{guri2015bitwhisper,masti2015thermal} discuss a thermal covert communication between different computers and cores via the control of heat emission and sensing. 

Sound waves, in the audible and non-audible bands, are also used to leak data from air-gapped systems \cite{madhavapeddy2005audio}. Managing `Audio modem' between laptops was proposed by Hanspach et al. Other works extended this method for malicious purposes where the attacker uses inaudible frequencies to establish covert mesh networking between laptops or workstations in a room \cite{hanspach2014covert} \cite{Guri2018Mosquito}. Researchers also showed that sound (sonic and ultrasonic) could be generated from a system with no audio hardware or speakers. They used various components to synthesize sound such as computer fans \cite{guri2020fansmitter}, HDD drives \cite{Guri2017f}, and power supplies \cite{9640597}.

\section{Technical Background}
\label{sec:tech}
SATA (Serial Advanced Technology Attachment) is a bus interface for connecting storage devices such as hard disk drives (HDD), Solid State Drives (SSD), and optical drives (CD/DVD) to a computer. SATA offers several advantages compared with the older standards such as PATA, such as larger bandwidths and higher transfer rates.

\begin{figure}
	\centering
	\includegraphics[width=0.8\linewidth]{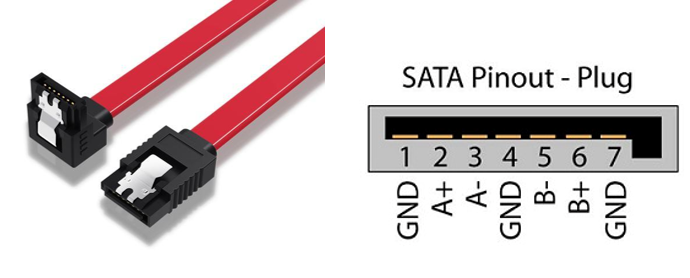}
	\caption{SATA cable and pinout scheme.}
	\label{fig:pinout}
\end{figure}

SATA cables are two side 7-pin cables, consisting of ground (pins 1,4,7), transmit (pins 2,3), and receive pins (pins 5,6) as specified in Table \ref{tab:SATAPIN} and Figure \ref{fig:pinout}. The ends are usually made at a 90-degree angle for better cable handling. The cable connects the SATA port on the motherboard and the storage device (e.g., HDD). There are smaller versions of SATA connectors, known as mSATA (mini-SATA), used with smaller computers, laptops, and tablets. The SATA power connector has supplies of  +3.3V DC, +5V DC, and +12V DC. SATA power cables that connect the device to the power supply unit (PSU) cables are often paired.

\begin{table}[]
	\centering
	\caption{SATA Data pinout}
	\label{tab:SATAPIN}
	\begin{tabular}{lll}
		\hline
		\textbf{Pin \#} & \textbf{Signal Name} & \textbf{Signal Description} \\ \hline
		\textbf{1}      & GND                  & Ground                      \\
		\textbf{2}      & A+                   & Transmit +                  \\
		\textbf{3}      & A-                   & Transmit -                  \\
		\textbf{4}      & GND                  & Ground                      \\
		\textbf{5}      & B-                   & Receive -                   \\
		\textbf{6}      & B+                   & Receive +                   \\
		\textbf{7}      & GND                  & Ground                      \\ \hline
	\end{tabular}
\end{table}

\begin{table}[]
	\centering
	\caption {SATA I,II,III transfer speeds}
		\label{tab:speed}
	\begin{tabular}{@{}lcc@{}}
		\toprule
		Standard & \multicolumn{1}{l}{Bandwidth} & \multicolumn{1}{l}{Data Transfer Speed} \\ \midrule
		SATA I   & 1.5 Gb/sec                    & 150 MB/sec                              \\
		SATA II  & 3 Gb/sec                      & 300 MB/sec                              \\
		SATA III & 6 Gb/sec                      & 600 MB/sec                              \\ \bottomrule
	\end{tabular}
\end{table}

\subsection{Transfer Rates}
The SATA standard has three main revisions that determine its bandwidth and data transfer speed, and other characteristics \cite{kawamoto2006hdd}. SATA revision 1.0, released in 2003, has a bandwidth of 1.5 Gbit/s and a data transfer speed of 150 MB/s. SATA revision 2.0, released in 2004, has a bandwidth of 3.0 Gbit/s and a data transfer speed of 300 MB/s. SATA revision 3.0, released in 2008, has a bandwidth of 6.0 Gbit/s and a data transfer speed of 600 MB/s. SATA 3 cables (third-generation SATA cables) are capable of transferring data at six gigabits per second. SATA 3 cables have locking latches on the ends of the cable. The three SATA major revisions, along with the bandwidth and transfer speeds, are presented in Table \ref{tab:speed}.  


\section{Transmission and Reception}
\label{sec:trans}
In this section, we describe the electromagnetic signal generation and the transmission protocol over the SATA interface. 

\subsection{Signal Generation}
The domain of information security of electromagnetic devices that are the source of unintentional emission induced in the surrounding area is known as TEMPEST \cite{NSTISSAM75:online}. When signals of undesirable emission are correlated with classified or confidential information, they can be used for reconstructing that information by intelligence entities. The phenomenon of such undesirable emissions is also known as the compromising emission threat. In the case of the SATAn covert channel, the attacker intentionally uses the SATA interface to generate and manipulate its electromagnetic emission. 

The SATA interface is a rich source of compromising radiated and conducted emission, mainly deriving from its data wires. Previous measurements of the radiated emission of the SATA 1.0 Shows that it spans a frequency range of 100 kHz to 1 GHz during the data transmission. In this work, we used the new SATA 3.0 interface to generate the emission \cite{przesrnvcki2018sata}. Notably, previous work shows that the transmitted data can not be intercepted since there is no relation between the \textit{content} of transmitted binary sequences and the compromising emission signals \cite{przesrnvcki2018sata}. Our experiments show a correlation between the bandwidth and data transfer speed and the radiated emission activity on the spectrum. More specifically, in SATA 3.0, the burst throughput of SATA 6.0 Gbit/s can be observed on the electromagnetic spectrum.

\begin{figure}
	\centering
	\includegraphics[width=1\linewidth]{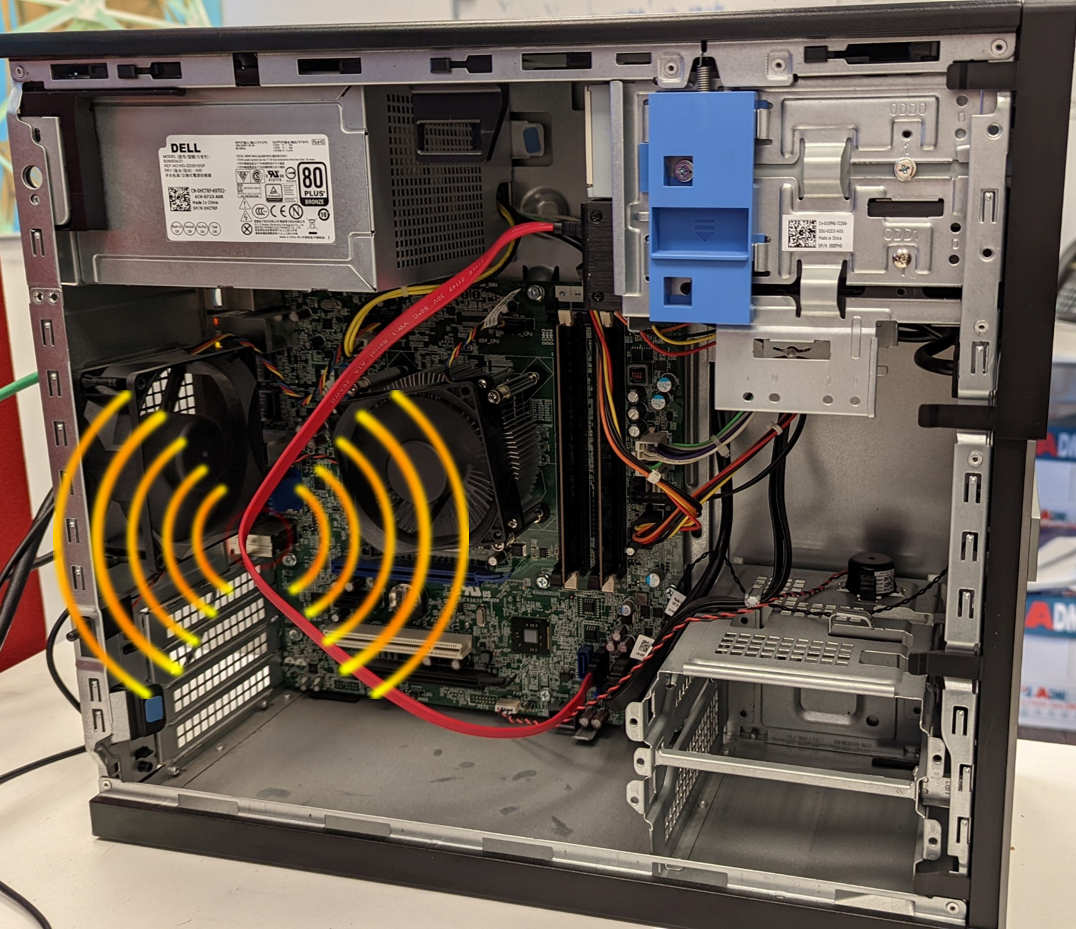}
	\caption{The SATA cable is used as an antenna to emanate electromagnetic signals (the case is open for the snapshot).}
	\label{fig:case}
\end{figure}

\begin{figure}
	\centering
	\includegraphics[width=1\linewidth]{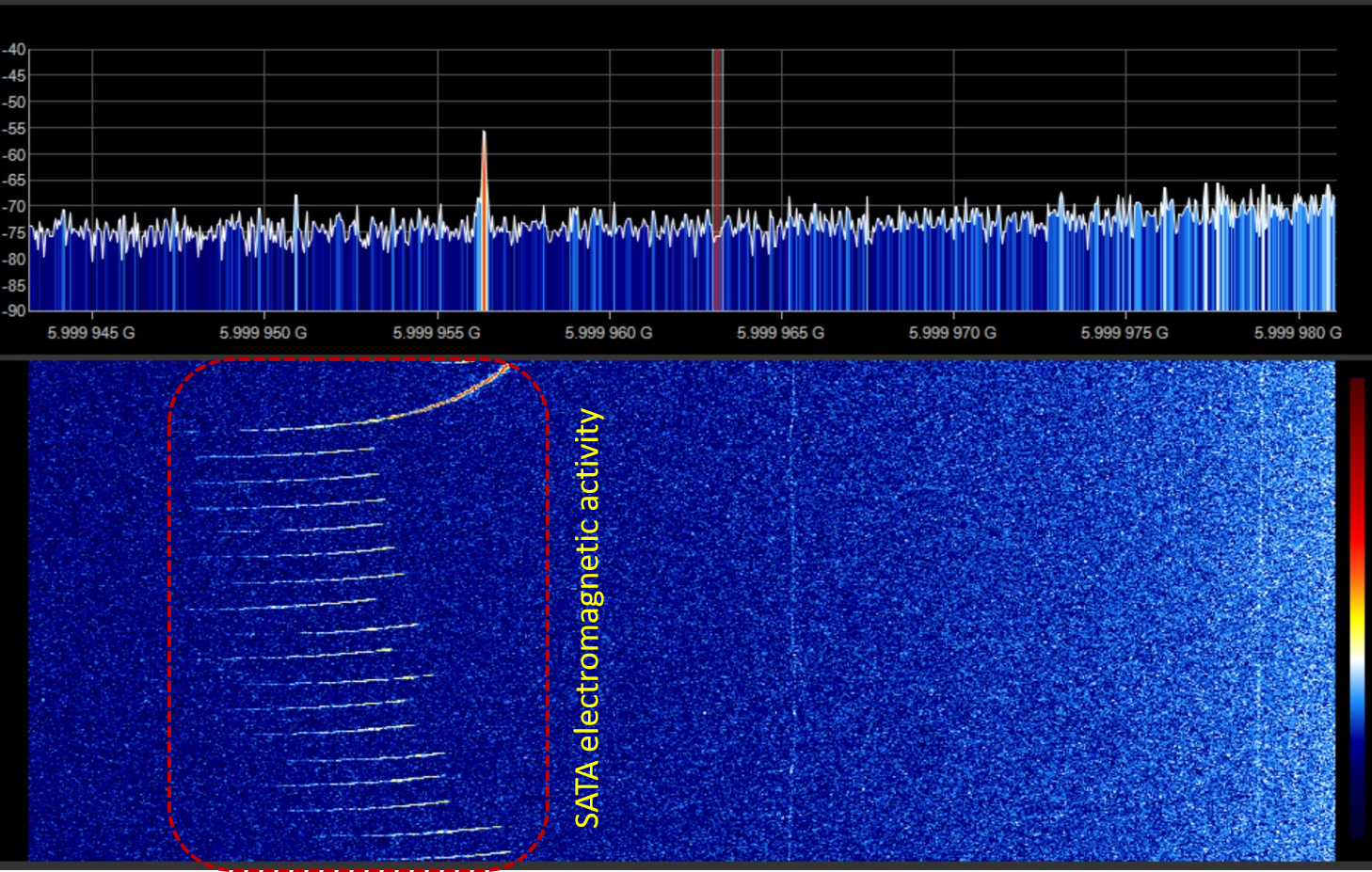}
	\caption{Electromagnetic emissions around 6 GHz emanated from the SATA interface due to a sequence of reading and writing operations.}
	\label{fig:EMR}
\end{figure}

Our experiments show that the SATA 3.0 cables emit electromagnetic emissions in various frequency bands; 1 GHz, 2.5 GHz, 3.9 GHz, and +\-6 GHz. However, the most significant correlation with the data transmission spans from 5.9995 GHz to 5.9996 GHz (Figure \ref{fig:EMR}). The idea behind the covert channel is to use the SATA cable as an antenna and control the electromagnetic emission, as shown in Figure \ref{fig:case}.

Algorithm \ref{alg0} shows the signal generation process. The version of the SATA-TRANSMIT function is generalized and can use both read and write operations of the transmission. 
The function receives the vector of bits to transmit (data), the bit-time for reading ($T_{read}$) and writing ($T_{write}$) operations,  and the percentage of times the read operation will be used from the whole operations ($S_{read}$). It also receives the time parameters for `0' ($T_{0}$) modulation. In the beginning, we disable the cache and buffering delay to make the SATA activity as instant as possible. Note that this might be achieved differently in different operating systems. For example, in the \texttt{open()} system call in Linux might receive the O\_DIRECT flag, which tries to minimize cache effects of the I/O to and from the file \cite{open2Lin93:online}. Other options are to use O\_SYNC or \texttt{fflush()} commands \cite{fflush3L62:online}. The function iterates on bits to transmit and randomize the current operation to perform (read or write) due to the user parameter, for $S_{read}$ and 1-$S_{write}$. 
The function reads or writes the data from a file or performs \texttt{sleep()} with the corresponding times, according to the current bit.

\begin{algorithm} 
	\caption{SATA-TRANSMIT (data, $T_{read}$, $T_{write}$, $S_{read}$ ,$T_{0}$)} 
	\label{alg0} 
	\begin{algorithmic}[1] 
		\State {$setNoCache(true)$}
		\State {$setNoDelay(true)$}
		\For {$i \gets 0\ to\ data.size()$}
		\State {$bit \gets data[i]$}
		
		\If {bit == 1}
		\State $currentOp \gets Random (T_{read}, T_{write}, S_{read}, 1-S_{read}) $ 
		
		\If {$currentOp == read $}
		\State $ read (datafile, T_{read}) $ 
		\EndIf
		\If {$currentOp == write $}
		\State $ open (tmpFile, "w") $ 
		\State $ randomData \gets randomBuff(1024) $ 
		\State $ write (tmpFile, randomData, T_{write}) $ 
		\EndIf
		\EndIf
		\If {$bit == 0$}
		\State $ sleep (T_{0}) $ 
		\EndIf
		
		\EndFor
		
	\end{algorithmic}
\end{algorithm} 

\subsection{Protocol}
Since the electromagnetic-based covert channel is unidirectional, we transmit the data in fixed-length data frames. Each frame begins with four alternating bits `1010' to broadcast the frame preamble. The frame preamble allows potential receivers to sync with the transmission. A 16-bit payload follows the preamble. The final information is a parity bit that is used for elementary error detection.

\subsection{AV Evasion}
\begin{figure}
	\centering
	\includegraphics[width=0.4\linewidth]{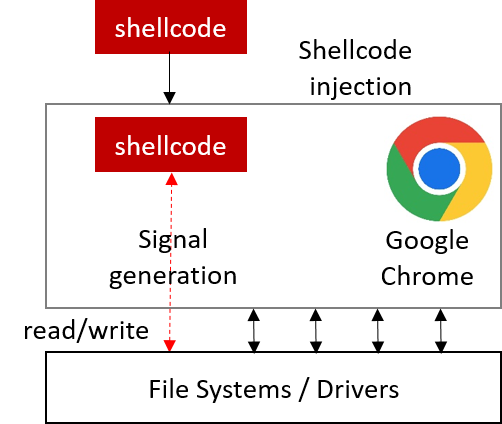}
	\caption{A signal generation shellcode was injected into the Google Chrome browser for evasion.}
	\label{fig:evasion}
\end{figure}

In order to evade anti-virus (AVs), Intrusion Detection Systems (IDS), and Intrusion Prevention Systems (IPS), the function can be implemented in a separate thread and injected into the memory space of another trusted process in the system. In Windows OS, such techniques are commonly used via the 'CreateRemoteThread', 'WriteProcessMemory', 'LoadLibrary', and similar APIs \cite{sikorski2012practical}. The advantage of the approach is that trusted processes in the system are allowed to write or read files in different folders frequently (e.g., temporary folders) without creating anomalies or being reported as suspicious activities (Figure \ref{fig:evasion}).   

\subsection{Receiver}
\label{sec:recv}
For the evaluation and testing purpose, we implemented the receiver in a MATLAB script. The reeving laptop is connected to a Software Defined Radio receiver and continently samples the output in the 5.9 GHz to 6 GHz frequency band. After performing the raw Fast Fourier Transform (FFT), the preamble is detected, and the payload is extracted. The pseudo-code for payload extraction is outlined in Algorithm \ref{alg1}.

\begin{algorithm} 
	\caption{SATA-DEMODULATE-FRAME (freq)} 
	\label{alg1} 
	\begin{algorithmic}[1] 
		\State {$setFreq(freq)$}
		\State {$delta \gets 1M$}
		
		\While {$find(vec, '1010') \neq true $}
		\State {$ vec += GetFFTVec(freq, delta)$}
		\EndWhile 
		
		\State {$T \gets preamble[1].time - preamble[0].time$}
		\While {$i < 16$}
		\State {$ vec = GetWindowedFFTVec(freq, delta, T)$}
		\If{$find (vec,'1')$} 
		\State {$payload += '1'$}
		\Else
		\State {$payload += '0'$}
		\EndIf
		\EndWhile 
		\State {$parity \gets extractParity()$}
		\State {$return (payload,parity)$}
		
	\end{algorithmic}
\end{algorithm} 

Note that in a real attack scenario, the receiver might be implemented as a process in the nearby computer or embedded in a dedicated hardware receiver.  

\begin{figure}
	\centering
	\includegraphics[width=0.8\linewidth]{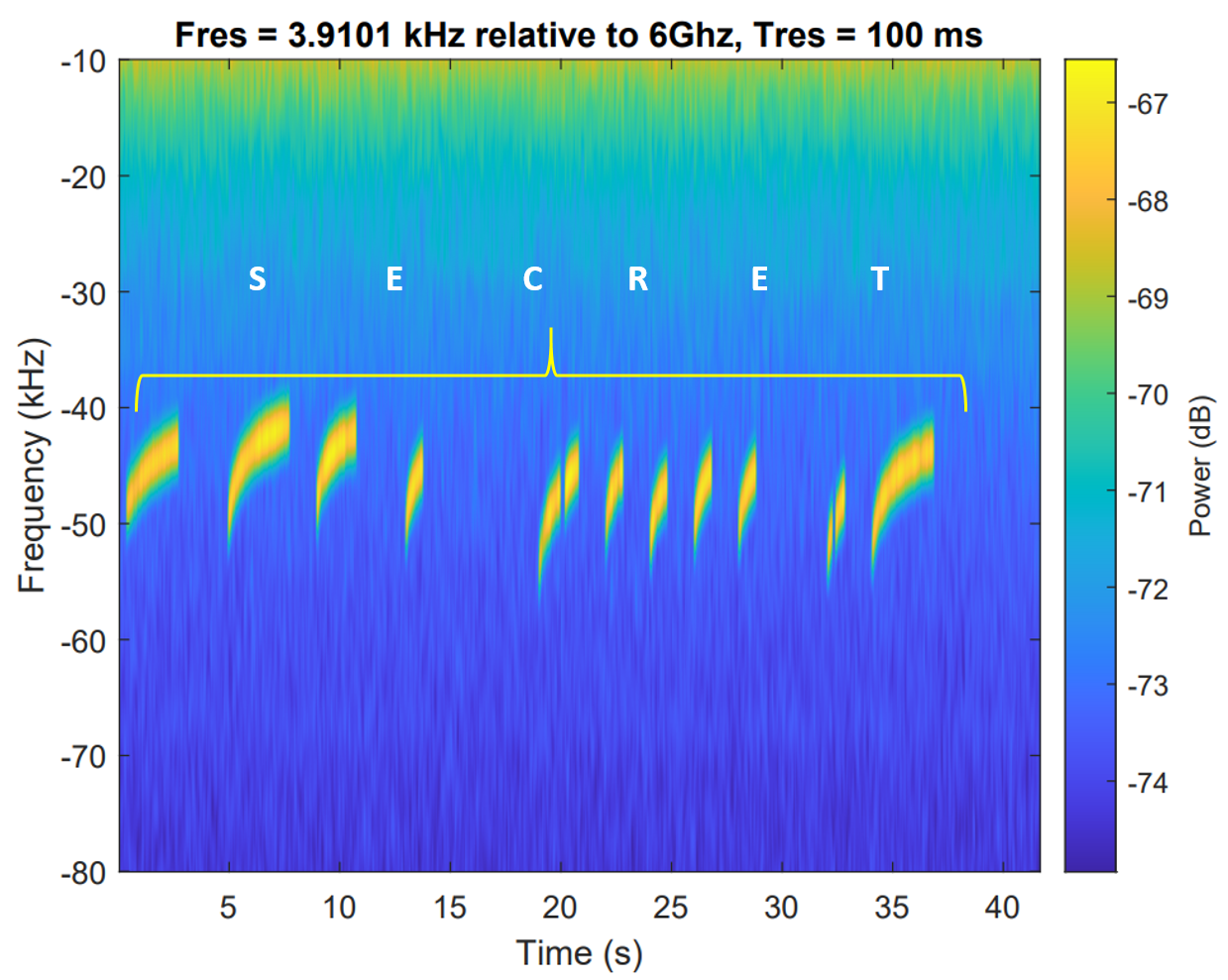}
	\caption{The payload 'SECRET' transmitted with the SATAn covert channel}
	\label{fig:sec}
\end{figure}

Figure \ref{fig:sec} shows the payload of the text 'SECRET' as transmitted by the covert channel. 

\section{Evaluation}
\label {sec:eval}
In this section, we present an evaluation of the SATAn covert channel. For the experimental setup, we used three off-the-shelf computers listed in Table \ref{tab:setup}. The computers under test have a metal chassis closed during the experiments. All tested computers had a SATA interface with Transcend 256GB MLC SATA III 6Gb/s 2.5" Solid State Drive 370 and were running Linux Ubuntu 20.04.1 64-bit. As a receiver, we used the ADALM PLUTO Software-defined Radio (SDR) AD9364 RF coverage from 70 MHz to 6 GHz. The SDR was connected through USB to a laptop with Microsoft Windows 10 Enterprise OS, and the output was processed by MathWorks MATLAB reception and demodulation script. 

\begin{table}[]
	\centering
	\caption {Experimental setup}
	\label {tab:setup}
	\begin{tabular}{@{}lll@{}}
		\toprule
		\#   & System                                                    & Operating System (OS)       \\ \midrule
		PC-1 & i7-4790 DELL 0N4YC8 8GB                                   & Linux Ubuntu 20.04.1 64-bit \\
		PC-2 & i7-6900k X99 ASRock 32GB & Linux Ubuntu 20.04.1 64-bit \\
		PC-3 & i3-4130 H81M-S2V 4GB                                      & Linux Ubuntu 20.04.1 64-bit \\ \bottomrule
	\end{tabular}
\end{table}

\subsection{Signal to Noise Ratio (SNR)}
Table \ref{tab:SNR} presents the signal-to-noise ratio (SNR) received with the three transmitting computers.
The signal transmitted from PC-1 has a strength of 20 dB at 30 cm to 9 dB at 120 cm apart. The signal generated from PC-1 and PC-2 were significantly weaker, with 15 dB at 60 cm (PC-2) and 7 dB at 30 cm (PC-3).       

\begin{table}[]
	\centering
	\caption {Signal to Noise Ratio (SNR)}
	\label {tab:SNR}
	\begin{tabular}{@{}lllll@{}}
		\toprule
		\#   & 30 cm                                & 60 cm                 & 90 cm                 & 120 cm                \\ \midrule
		PC-1 & 10 dB / 20 dB                        & 7 dB / 15 dB          & 6 dB / 13 dB          & 4 dB / 9 dB           \\
		PC-2 & 7 dB / 1 dB & 3 dB/ 9 dB            & \multicolumn{1}{c}{-} & \multicolumn{1}{c}{-} \\
		PC-3 & 3 dB / 7 dB                          & \multicolumn{1}{c}{-} & \multicolumn{1}{c}{-} & \multicolumn{1}{c}{-} \\ \bottomrule
	\end{tabular}
\end{table}

\subsection{Bit Times}
\begin{figure}
	\centering
	\includegraphics[width=0.8\linewidth]{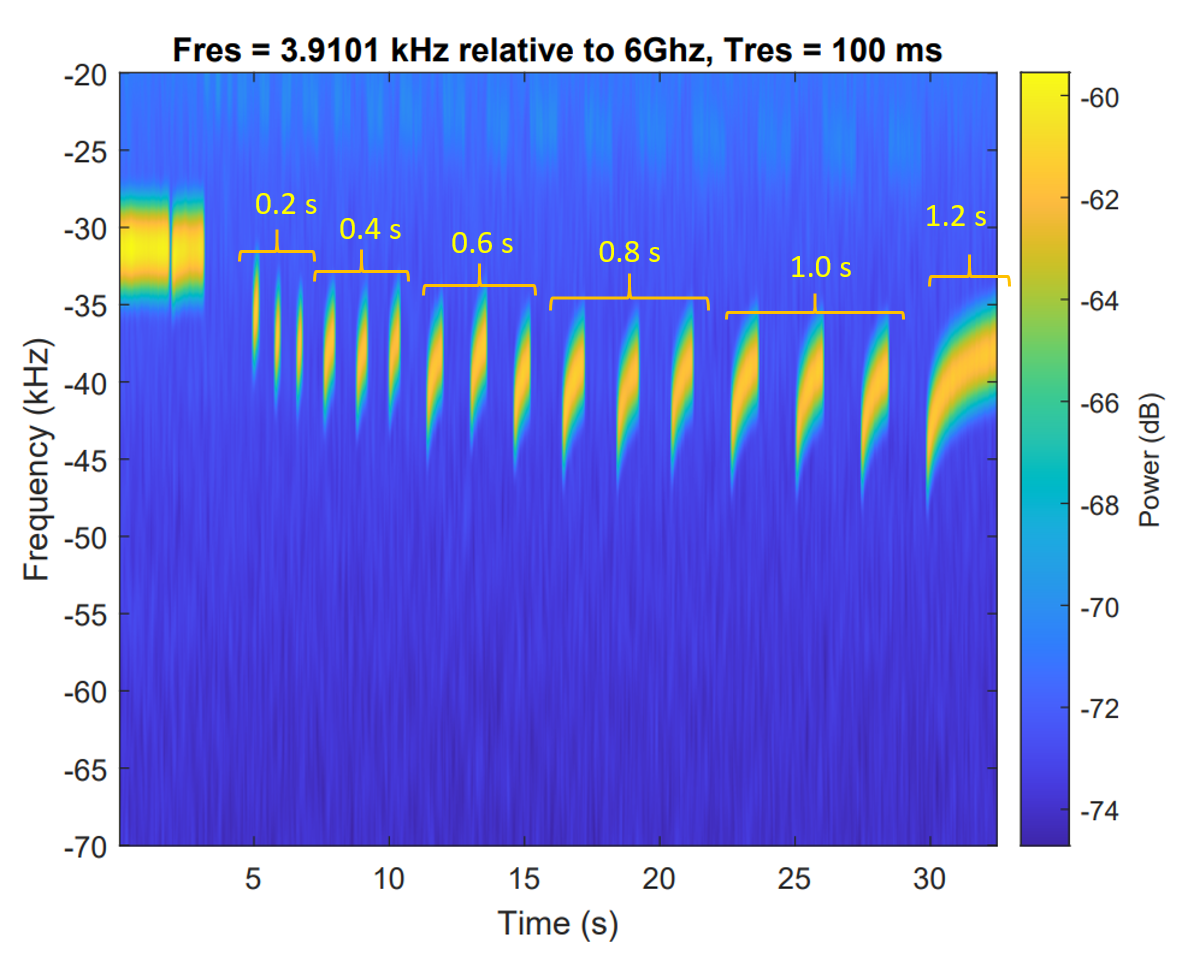}
	\caption{Signal generated with timing of 0.2, 0.4, 0.6, 0.8, 1.0, and 1.2 seconds.}
	\label{fig:speed}
\end{figure}

The timing parameter of the read and write operations $T_{read}$ and $T_{write}$ has a direct effect on the SATA activity, the electromagnetic emission, and the generated signal time. Since we use On-off-keying for modulation, the `1' and '0` times $T_{1}$ and $T_{0}$ correlate with the read/write operation times. 
Figure \ref{fig:speed} shows the spectrogram of a bit transmission with different timing parameters. In this case sequences of three bits with 0.2 sec, 0.4 sec, 0.6 sec, 0.8 sec, 1.0 sec, and 1.2 sec have been modulated and received. The SNR levels with the different times are given in Table \ref{tab:SNRBIT}. 

\begin{table}[]
	\centering
	\caption {Bit times and the corresponding signal}
	\label {tab:SNRBIT}
	\begin{tabular}{@{}lllllll@{}}
		\toprule
		T      & 0.2 sec & 0.4 sec & 0.6 sec & 0.8 sec & 1.0 sec & 1.2 sec \\ \midrule
		Signal &     -72 dBm    &    -72 dBm      &    -70 dBm      &    -70 dBm     &     -68 dBm    &    -68 dBm     \\ \bottomrule
	\end{tabular}
\end{table}

\subsection{Bit Error Rate (BER)}
We transmitted the data with a bit rate of 1 bit/sec, which is shown to be the minimal time to generate a signal which is strong enough for modulation.
The BER for PC-1 is presented in Table \ref{tab:BER}. As can be seen, the BER of 1\% - 5\% is maintained between 0 - 90 cm. With a greater distance of 120 cm, the BER is significantly higher and reaches 15\%.
With PC-2 and PC-3, the bit error rates (BER) are less than 5\% only in short proximity up to 30 cm, and hence the attack is relevant only for short ranges in these computers.     

\begin{table}[]
	\centering
	\caption {PC-1 BER}
	\label {tab:BER}
	\begin{tabular}{@{}lllll@{}}
		\toprule
		\#   & 30 cm & 60 cm & 90 cm & 120 cm \\ \midrule
		PC-1 & 1\%   & 3\%   & 5\%   & 15\%   \\ \bottomrule
	\end{tabular}
\end{table}

\subsection{Read vs. Write Operations}
\begin{figure}
	\centering
	\includegraphics[width=1\linewidth]{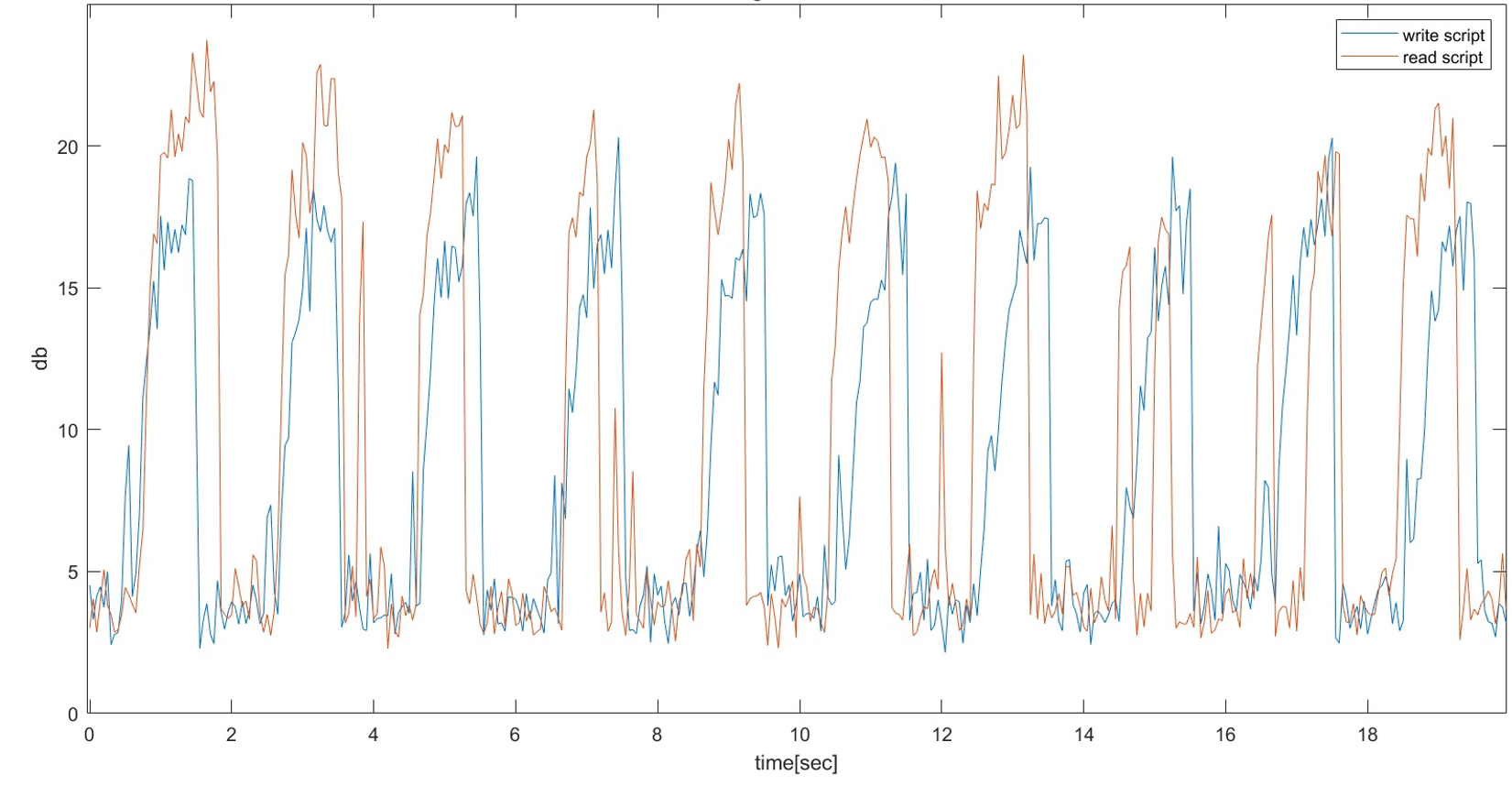}
	\caption{SNR of read and write operations.}
	\label{fig:RW}
\end{figure}

The signal can be generated with reading or writing operations and translated into the corresponding ATA read or write commands at the hardware level. Figure \ref{fig:RW} presents the signal generated with reading and writing operations, where a sequence of alternating bits was transmitted from PC-1. The results show that read operations yield a signal with an average of 3 dB stronger than write operations. It means that it is preferable to use read operation for the covert channel. Notably, read operations may require lower permissions than write operations. For example, an application may be permitted to read data or configuration files but might be restricted in writing to them.  

\subsection{Background Processes}
Background processes in the system are generating electromagnetic emissions due to the internal activity in the controller, buses, and motherboard computers. We checked the effect of typical CPU-bound and I/O-bound workloads on the signal's quality. Table \ref{tab:workloads} presents the different workloads and the SNR measured with these activities. In an idle state where no special processes run in the background, the signal generation yields an SNR of 15 dB - 20 dB. Our test shows that the signal generation in an idle state where no special processes run in the background yield an SNR of 15 dB - 20 dB. Intensive computation and virtual memory operations didn't affect the signal quality and yielded an SNR of 15 dB - 20 dB. The daily workload of an interactive process such as word processing and network bound such as YouTube video playing in Google Chrome web browser didn't affect the signal quality either and yielded an SNR of 15 dB - 20 dB. The intensive disk operations with file transfer cause significant signal degradation to 4.5 dB - 5.5 dB. The above results indicate that the attack can be a maintained event with active workloads on the system, which are CPU and I/O bound. However, the covert channel is rendered less effective when intensive disk activity is involved due to the reduced quality. From the attacker's point of view, where intensive disk activity is detected, the transmission should be halted or postponed to a later time.

\begin{table}[]
	\centering
	\caption {Signal quality with different workload}
	\label {tab:workloads}
	\begin{tabular}{@{}llll@{}}
		\toprule
		\# & Activity                                & Type                  & SNR             \\ \midrule
		1  & Idle                                    & \multicolumn{1}{c}{-} & 15 dB - 20 dB   \\
		2  & Intensive computation (100\% CPU)       & CPU-bound             & 15 dB - 20 dB   \\
		3  & Intensive RAM activity & CPU-bound             & 15 dB - 20 dB   \\
		4  & Word processing application             & Interactive           & 15 dB - 20 dB   \\
		5  & YouTube video playing                   & I/O  (network)        & 15 dB - 20 dB   \\
		6  & Copy files between directories          & I/O (disk)            & 4.5 dB - 5.5 dB \\ \bottomrule
	\end{tabular}
\end{table}

\subsection{Virtual Machine (VM)}
\begin{figure}
	\centering
	\includegraphics[width=1\linewidth]{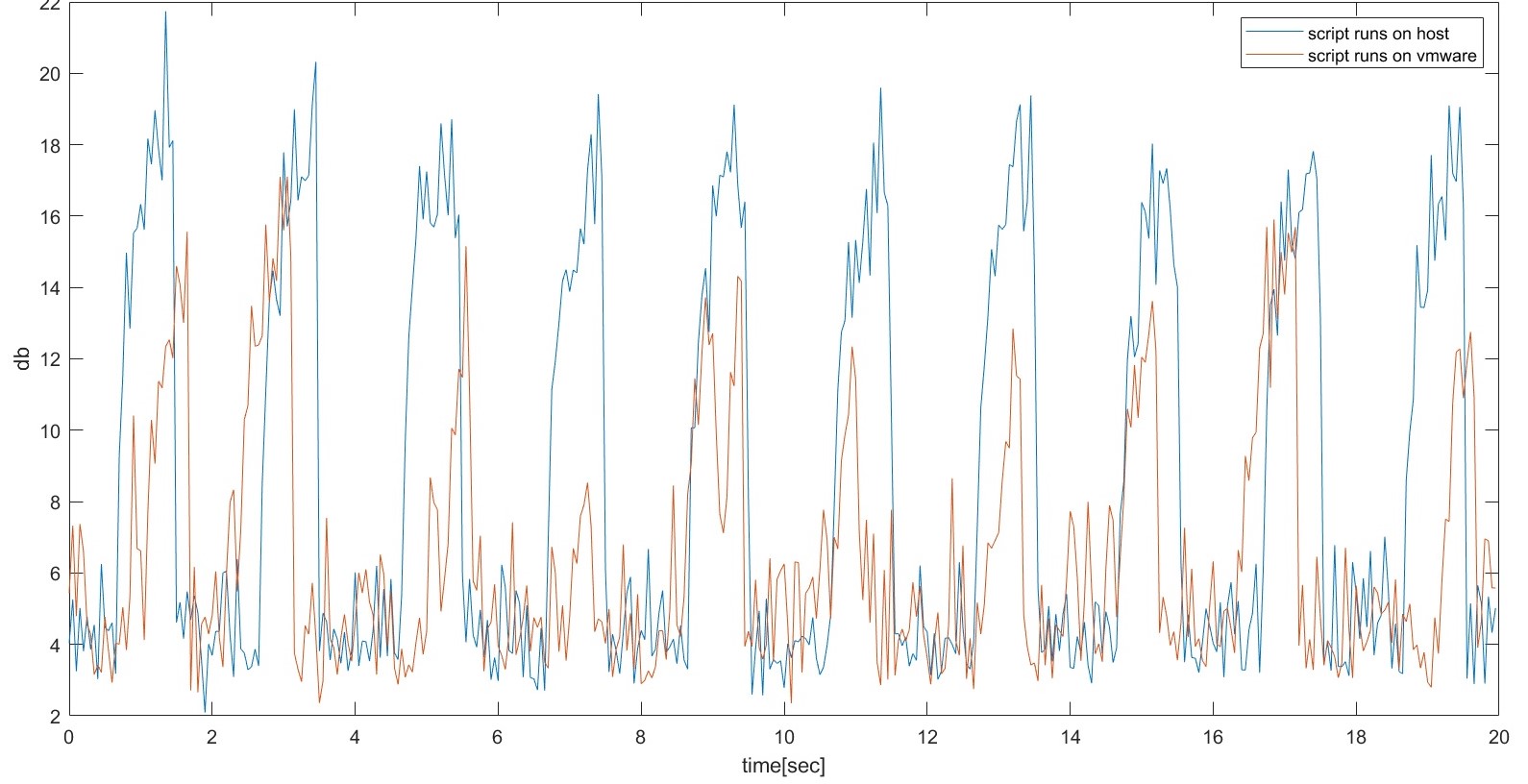}
	\caption{VMM vs. host signal generation.}
	\label{fig:VMM}
\end{figure}

Virtual Machines (VM) are widely used technology in modern working environments, including in servers, local networks, cloud environments, and personal workstations. The Virtual Machine Monitor (VMM) or hypervisor creates a layer of abstraction where a physical host is virtualized at the hardware of the operating system level, enabling multiple isolated and secure virtualized guests to run on a single physical machine. The VMM usually provides host-level storage virtualization, which logically abstracts the physical storage layer from virtual machines. The VM uses a virtual disk to store its operating system, program files, and other data associated with its activities. Many VMMs, such as Oracle's VirtualBox, Microsoft's Virtual PC, and VMware use a single-file virtual file system to manage the virtual storage device \cite{hansen2010lithium}. In the context of our covert channel, it means that the read/write commands are not necessarily sent directly through the SATA interface but may through layers of buffering and caching in the VMM and the host, which may cause a delay or interruption to the generated signal. We tested the signal generation from inside a VMWare VM with the six workloads from table \ref{tab:workloads}. The results show an average reduction of  5 dB in the signal quality in VM compared to the signal generated from the host, as shown in Figure \ref{fig:VMM}. These differences are due to the inconsistent read/write activity in the physical SATA interface when the read/write operations are executed from within a VM.

\section{Countermeasures}
\label{sec:counter}
\begin{figure*}
	\centering
	\includegraphics[width=0.8\linewidth]{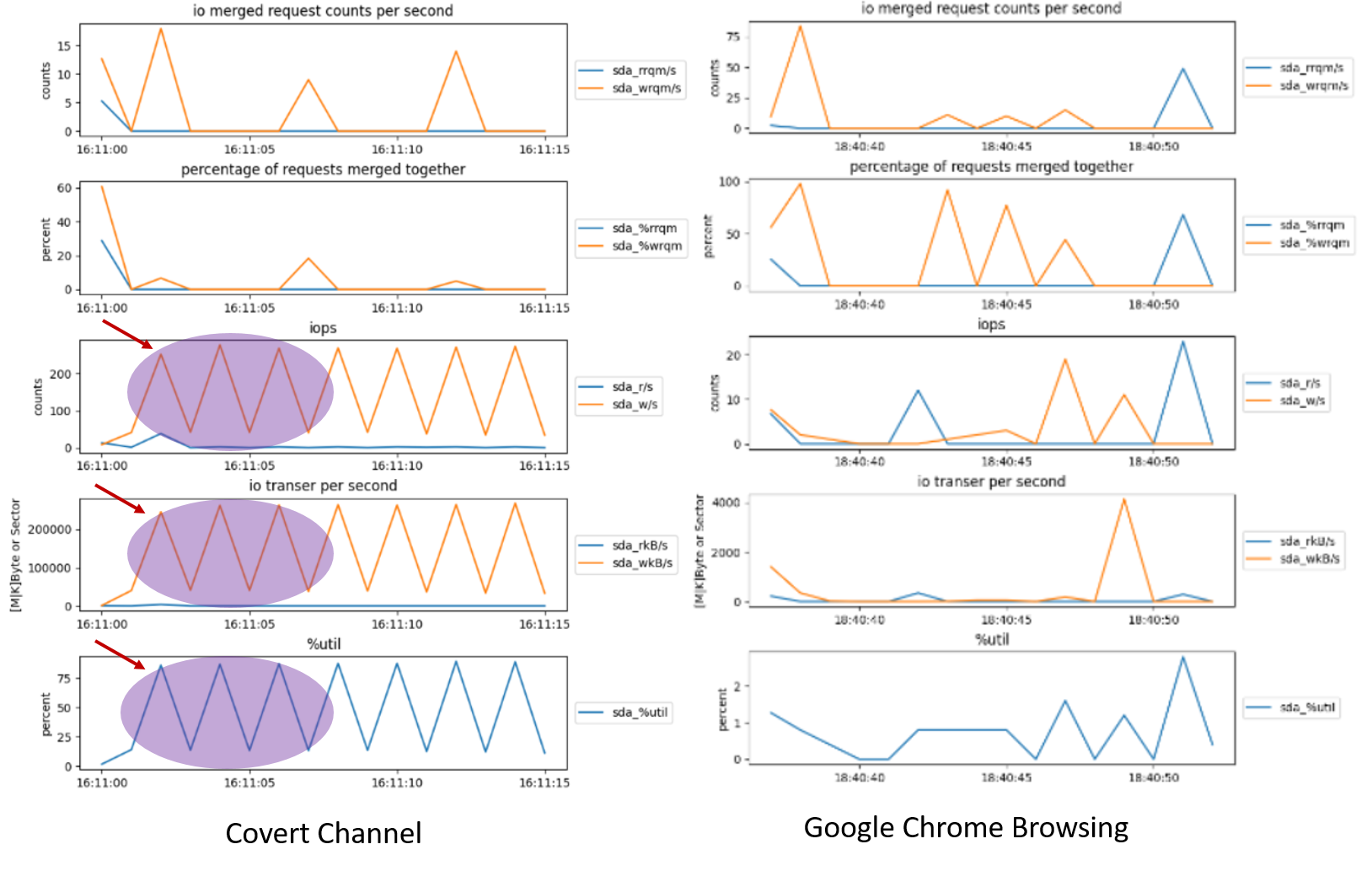}
	\caption{Detection of the covert channel (left) using measurements of I/O transfer per second and disk utilization over time.}
	\label{fig:detect}
\end{figure*}

There are different classes of countermeasures for the electromagnetic covert channels. Preventing the initial penetration is the first step that should be taken as a preventive countermeasure. There is a wide range of network security technologies that protect the usability and integrity of a company's infrastructure. To prevent the first phase of the air-gap penetration, multiple layers of security should be used in the network, including firewalls, intrusion detection and prevention systems, network traffic analysis, and access control mechanisms. Policy-based countermeasures might forbid the use of radio receivers in secured facilities or rooms within a certain distance. This approach is also known as 'red/black` separation and is mentioned by US and NATO standards \cite{NSTISSAM75:online}. Another approach is to use an external RF monitoring system to detect anomalies in the 6 GHz nearby the transmitting computer. However, this approach will likely suffer from false alarms and a low detection rate since any read/write operation would create electromagnetic emission in this range, regardless of the covert channel. In order to detect the signal, the system has to know the specific signal shape and modulation in use, which is less practical from the defender's perspective. It is possible to install a dedicated driver (e.g., filter driver in Windows OS) that detect abnormal read/write operations. In the case of this covert channel, anomalous read and write operations from or to temporary files would trigger alerts. Figure \ref{fig:detect} shows our anomaly detection of the covert channel by monitoring the I/O operations per second and the disk utilization of a process. The anomalous pattern of the transmitting process can be clearly observed in the covert channel process (left) compared to a Google Chrome browsing (right). However, its important to note the detection of such dummy operations is highly contextual to the specific process and very challenging in a runtime environment, mainly due to the inability to distinguish between legitimate and malicious operations.

 Other preventing types of countermeasure belong to the jamming category. The jamming can be done from the operating system by performing random read and write operations when a suspicious covert channel activity is detected. Algorithm \ref{alg3} shows the outline of the jammer function SATA-JAM. The function received the parameters for a maximum bit time ($T$). A random read or write operation is initiated for a random time of (0..$T$).

\begin{algorithm} 
	\caption{SATA-JAM (T)} 
	\label{alg3} 
	\begin{algorithmic}[1] 
		
		\For {$i \gets 0\ to\ data.size()$}
		\State {$tm \gets Random (T)$}
		
		\State $currentOp \gets Random (0.5, 0,5, read, write) $ 
		
		\If {$currentOp == read $}
		\State $ read (datafile, tm) $ 
		\EndIf
		\If {$currentOp == write $}
		\State $ open (tmpFile, "w") $ 
		\State $ randomData \gets randomBuff(1024) $ 
		\State $ write (tmpFile, randomData, tm) $ 
		\EndIf
		
		\EndFor
		
	\end{algorithmic}
\end{algorithm} 

\begin{figure}
	\centering
	\includegraphics[width=1\linewidth]{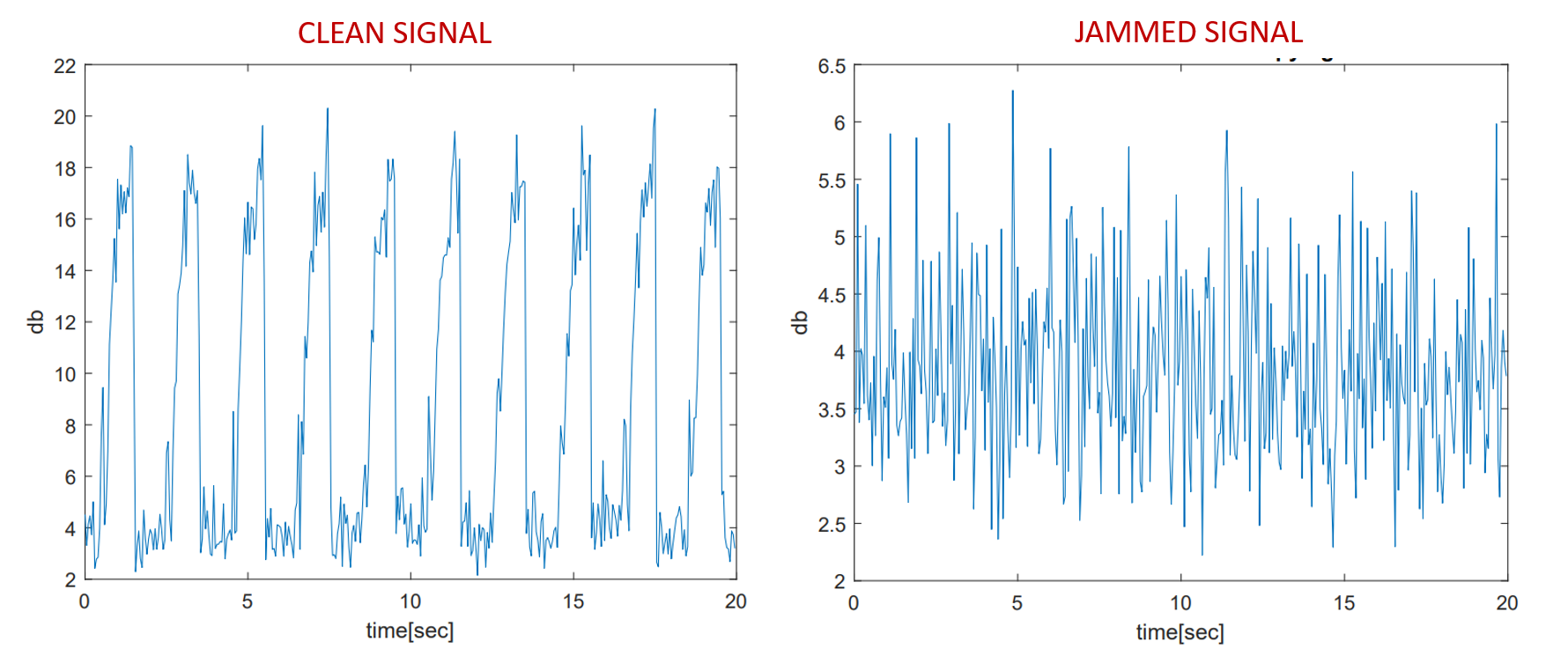}
	\caption{A clean signal and jammed signal.}
	\label{fig:jam}
\end{figure}

Figure \ref{fig:jam} shows the power of a clean signal and a jammed signal. The signal was jammed for over 20 seconds using the random read and write operations. As observed, the SNR of the jammed signal is significantly reduced to  3.5 dB - 5.5 dB. 

As observed in Section \ref{sec:eval}, the existence of intensive disk operations causes an interruption to the signal generation process and reduces the quality of the signal. However, this solution has a significant drawback of harming the performance of disk and I/O activities in the OS and, in the long term, may cause damage to the storage. The external jamming approach would involve the use of radio signal jammers in the 6 GHz frequency band. However, such devices are expensive and can not be practically deployed on a wide scale.

\section{Conclusion}
\label{sec:conclusion}
This paper presents SATAn - a new type of attack on air-gapped computers. We show that attackers can exploit the SATA cable as an antenna to transfer radio signals in the 6 GHz frequency band by using non-privileged read() and write() operations. Notably, the SATA interface is highly available to attackers in many computers, devices, and networking environments. We discuss related work and provide technical background. We show the design of the covert channel and present the implementation of the transmitter and receiver. The results show that attackers can use the SATA cable to transfer a brief amount of sensitive information from highly secured, air-gap computers wirelessly to a nearby receiver more than 1m away. We also show that the attack can operate from user mode and is effective even from inside a guest VM. We also discuss preventive and protective countermeasures to this covert channel attack.

\def\UrlBreaks{\do\/\do-}
\balance
\bibliographystyle{plain}
\bibliography{SATA,../../AirGap,../../AirGapCases,../../mobile,../../AirGapTools,../../PowerSupply}
\end{document}